\newif\ifarticle
\newif\ifaps
\newif\ifwordcount

\apstrue 

\ifarticle
\documentclass[a4paper,11pt]{article}
\fi

\ifaps
\ifwordcount
\documentclass[aps,prd,reprint,nofootinbib]{revtex4-1}
\else
\documentclass[aps,prd,reprint]{revtex4-1}
\fi
\fi

\usepackage[utf8]{inputenc}
\usepackage[T1]{fontenc}
\usepackage{lmodern}
\usepackage{microtype}
\usepackage{amsmath,amsfonts,amssymb,amsthm}
\usepackage{mathrsfs}
\usepackage[hidelinks]{hyperref}
\usepackage{cleveref}
\usepackage{graphicx}
\usepackage{tensor}
\usepackage{braket}
\usepackage{comment}
\usepackage{array}
\usepackage{booktabs}
\usepackage{mathtools}

\ifarticle
\usepackage{authblk}
\usepackage{cite}
\newcommand{\authorAndEmail}[2]{\author{#1\thanks{\href{mailto:#2}{#2}}}}
\newenvironment{acknowledgments}{\section*{Acknowledgments}}{}
\let\affiliation\affil
\date{\vspace{-3em}}
\let\oldbibliography\thebibliography
\renewcommand{\thebibliography}[1]{\oldbibliography{#1}\setlength{\itemsep}{0pt}}
\fi

\ifwordcount
\usepackage[disable]{todonotes}
\else
\usepackage{todonotes}
\fi

\ifaps
\newcommand{\authorAndEmail}[2]{\author{#1}\email{#2}}

\ifwordcount
\usepackage{environ}
\NewEnviron{envhidden}{}

\fi
\fi

\crefname{equation}{}{}

\newcommand*{\dd}{\mathop{}\!d}

\newcommand*{\pd}{\mathop{}\!\partial}

\newcommand*{\scri}{\ensuremath{\mathscr{I}}}
\newcommand*{\BMS}{\ensuremath{\textrm{BMS}}}

\begin{document}

\title{Linking Past and Future Null Infinity in Three Dimensions}
\authorAndEmail{Stefan Prohazka}{prohazka@hep.itp.tuwien.ac.at}
\authorAndEmail{Jakob Salzer}{salzer@hep.itp.tuwien.ac.at}
\authorAndEmail{Friedrich Schöller}{schoeller@hep.itp.tuwien.ac.at}
\affiliation{Institut für Theoretische Physik, Technische Universität Wien \\
  Wiedner Hauptstraße 8--10/136, 1040 Wien, Austria
}

\ifarticle\maketitle\fi

\begin{abstract}
We provide a mapping between past null and future null infinity in three-dimensional flat space, using symmetry considerations.
From this we derive a mapping between the corresponding asymptotic symmetry groups.
By studying the metric at asymptotic regions, we find that the mapping is energy preserving and yields an infinite number of conservation laws.
\end{abstract}

\ifaps
\ifwordcount\else\maketitle\fi
\fi

\section{Introduction}
\label{sec:introduction}

Three-dimensional theories have a long history as toy models in quantum gravity.
Often they allow for calculations currently out of reach in higher dimensions, and provide insights into deep conceptual problems.
Recently, the rich infrared structure of perturbative quantum gravity in four-dimensional asymptotically flat spacetimes has attracted increased attention.
The asymptotic boundary of these spacetimes contains past and future null infinity denoted by $\scri^-$ and $\scri^+$, respectively.
Both 
are separately invariant under an infinite-dimensional symmetry group, the Bondi-Metzner-Sachs (BMS) group~\cite{Bondi:1962px,Sachs:1962zza}. Surprisingly, this symmetry group is intimately related to both the gravitational memory effect and Weinberg's soft graviton theorem~\cite{Strominger:2013jfa,Strominger:2014pwa,He:2014laa}.
In particular, the latter arises as a Ward identity for BMS invariance of the S-matrix.
To consider the BMS group as a symmetry of the S-matrix one must relate the two --- a priori independent --- symmetry groups at each boundary.

In this work we propose a linking between the two asymptotic regions and their symmetries in three-dimensional Einstein gravity.

In four and higher, even dimensions, this was accomplished previously~\cite{Strominger:2013jfa,Kapec:2015vwa} (although for the higher-dimensional case see the objections~\cite{Hollands:2016oma}).
In the present work we cover what seems to be the only remaining case of physical interest.
The framework of conformal null infinity does not appear to be useful in odd spacetime dimensions higher than three~\cite{Hollands:2004ac}.

Three-dimensional pure Einstein gravity does not exhibit local degrees of freedom, i.e., gravitational waves, but the theory possesses degrees of freedom on the boundary.
Nontrivial scattering in the interior is obtained by coupling the theory to propagating matter.
Due to its technical simplicity, e.g., detailed knowledge of the phase space, the theory then provides a unique testing lab for further studies of the infrared sector of quantum gravity, building upon~\cite{He:2014laa,Strominger:2013jfa}.
We provide a first step toward studying such a setup and its relation to BMS symmetry by breaking the two separate BMS symmetries, ending up with a single global one.

Attempts at a holographic framework of asymptotically flat spacetimes yield another motivation for our work.
Compared to anti-de Sitter (AdS) space, where holography is realized in form of the Anti-de Sitter/conformal field theory (AdS/CFT) correspondence, flat space holography is still poorly understood.
AdS$_3$/CFT$_2$ is one of the prime examples of holography, due to the high level of control over both sides of the correspondence.
Given the conceptual clarity of AdS holography in three dimensions, three-dimensional space suggests itself as a natural testing ground for ideas of flat space holography.

Most of the recent evidence~\cite{%
Bagchi:2012yk,Barnich:2012xq,Bagchi:2012xr,Barnich:2012rz,
Barnich:2013yka,Bagchi:2013lma,Fareghbal:2013ifa,
Detournay:2014fva,Riegler:2014bia,Fareghbal:2014qga,Bagchi:2014iea,
Barnich:2015mui,Garbarz:2015lua,Bagchi:2015wna,Hartong:2015usd,Bonzom:2015ans,Basu:2015evh,Campoleoni:2015qrh,
Campoleoni:2016vsh,Carlip:2016lnw,Bagchi:2016bcd,Bagchi:2016geg,Asadi:2016plj%
} for a field theory dual to Einstein gravity on three-dimensional flat space was focused on one connected component of $\scri$ only.
A holographic framework for flat spacetimes should benefit from considerations involving both null boundary components.

In  \cref{sec:flat} we start by providing boundary conditions, asymptotic symmetries and charges for our spacetimes.
Following a discussion of the phase space of vacuum solutions in  \cref{sec:phase-space} we provide a linking of their asymptotic regions in \cref{sec:linking} using symmetry arguments.
In \cref{sec:matter} we argue that the linking can be generalized to hold when matter is present.

\section{Asymptotically Flat Spacetimes}
\label{sec:flat}


Asymptotically flat spacetimes at future (past) null infinity are spacetimes that admit a conformal null-boundary $\scri^+$ ($\scri^-)$ in the future (past)~\cite{Penrose:1962ij}.
Equivalently, they are spacetimes such that the metric, by a suitable choice of coordinates, can be brought into the form (cf.~\cite{Tamburino:1966zz} in four dimensions)
\begin{align}
\label{eq:1}
  \dd s^2
  &= r^{-1} V^+ e^{2\beta^+} \dd u^2
    - 2e^{2\beta^+}\dd u \dd r
    + r^2(\dd \phi - U^+\dd u)^2
\end{align}
around $\scri^+$ and similarly around $\scri^-$,
\begin{align}
  \label{eq:2}
  \dd s^2
  &= r^{-1} V^- e^{2\beta^-} \dd v^2
    + 2e^{2\beta^-}\dd v \dd r
    + r^2(\dd \phi - U^-\dd v)^2
    \, ,
\end{align}
where $\phi \sim \phi + 2 \pi$.
The functions $U^\pm$, $V^\pm$ and $\beta^\pm$ depend on $u$, $r$ and $\phi$, and satisfy
\begin{equation}
  \lim_{r \to \infty} U^\pm
  = \lim_{r \to \infty} \beta^\pm
  = \lim_{r \to \infty} r^{-3} V^\pm
  = 0
  \,.
\end{equation}
Here $u$ and $v$ are retarded and advanced time coordinates.

Diffeomorphisms preserving the form of the metric act as
\begin{align}
\label{eq:bms-diffeos}
\begin{aligned}
  u
  &\to u f'(\phi) + \alpha(f(\phi)) + O(r^{-1})
  \\
  r
  &\to r / f'(\phi) + O(1)
  \\
  \phi
  &\to f(\phi) + O(r^{-1})
    \, ,
\end{aligned}
\end{align}
around $\scri^+$ and similarly around $\scri^-$.
The function $f$ is required to be a diffeomorphism on the circle and parametrizes so-called superrotations, which generalize Lorentz transformations. Translations are generalized to the so called supertranslations $\alpha$.
Together they form the asymptotic symmetry group, the three-dimensional BMS group~\cite{Bondi:1962px,Sachs:1962zza,Ashtekar:1996cd}.

In four dimensions the BMS group, originally presented in~\cite{Bondi:1962px,Sachs:1962zza}, is the semidirect product of globally well-defined conformal transformations of the sphere, i.e., the Lorentz group, and the infinite-dimensional abelian group of supertranslations.
Recently, it was proposed to allow for conformal transformation of the sphere that are well-defined only locally, called superrotations~\cite{Barnich:2009se,Banks:2003vp} or to allow for arbitrary diffeomorphisms of the sphere~\cite{Campiglia:2014yka}.
In three dimensions, two of the three options coincide, since all diffeomorphisms of the circle are also conformal transformations.
Here, the superrotations have the group structure of Diff($S^1$) and are, in contrast to the higher dimensional case, globally well defined.

Diffeomorphisms that are restricted to the bulk of spacetime are proper gauge transformations, so the diffeomorphisms~\cref{eq:bms-diffeos} can be continued arbitrarily into the bulk.
Moreover, their form around $\scri^+$ is a priori not related to their form around $\scri^-$.
It follows that there is the freedom of choosing the coordinate systems~\cref{eq:1} and~\cref{eq:2} independently.
This freedom is precisely expressed by the BMS group acting on $\scri^+$, which we refer to as $\mathrm{BMS}^+$
and the one acting on $\scri^-$, $\mathrm{BMS}^-$.

Metrics of the form~\cref{eq:1} and~\cref{eq:2}, solving the vacuum Einstein equations, have the remarkably simple form~\cite{Barnich:2010eb,Barnich:2012aw}
\begin{multline}
  \label{eq:3}
  \dd s^2
  = \Theta^+ \dd u^2
  - 2\dd u \dd r
  + \left(2 \Xi^+ + u \pd_\phi \Theta^+ \right) \dd u \dd \phi
  +{} \\
  + r^2 \dd \phi^2
  \,,
\end{multline}
and
\begin{multline}
  \label{eq:7}
  \dd s^2
  = \Theta^- \dd v^2
  + 2\dd v \dd r
  + \left(2 \Xi^- + v \pd_\phi \Theta^- \right) \dd v \dd \phi
  +{} \\
  + r^2 \dd \phi^2
  \,,
\end{multline}
with arbitrary functions $\Theta^\pm(\phi)$ and $\Xi^\pm(\phi)$.
They are called  mass aspect and angular momentum aspect, respectively.

The charges associated to the symmetries~\cref{eq:bms-diffeos} were calculated~\cite{Barnich:2006av} using covariant phase space methods~\cite{Barnich:2001jy}.
They are given by
\begin{align}
\label{eq:charges}
  Q_{T,Y}
  &= \frac{1}{16 \pi G} \int_0^{2\pi} \left(\Theta T + 2 \Xi Y \right) \dd\phi
    \,,
\end{align}
where $T(\phi)$ and $Y(\phi)$ parametrize infinitesimal supertranslations and superrotations, respectively.
This shows that spacetimes with different values of $\Theta$ and $\Xi$ can be distinguished by their charges.
The energy of a spacetime is given by the charge $Q_{1,0}$, its angular momentum by $Q_{0,1}$.

Under a finite BMS transformation~\cref{eq:bms-diffeos}, the functions $\Theta$ and $\Xi$ transform as~\cite{Barnich:2012rz}
\begin{align}
\label{eq:finBMS}
\begin{aligned}
  \Theta
  &\to
    (f')^2 \Theta \circ f - 2 S[f]
  \\
  \Xi
  &\to
    (f')^2 \Big[
    \Xi
    + \frac{1}{2} \Theta' \alpha
    + \alpha' \Theta
    - \alpha'''
    \Big] \circ f
   \,,
\end{aligned}
\end{align}
where $S[f]$ denotes the Schwarzian derivative.
Transformations not changing $\Theta$, and thus preserving the energy, create soft gravitational modes.

In the following sections we derive a mapping between the two asymptotic regions, which then leads to the linking of the symmetry groups $\BMS^+$ and $\BMS^-$.

\section{Phase Space and Validity of the Mapping}
\label{sec:phase-space}

In this section we collect results on the phase space of three-dimensional, asymptotically flat gravity without matter 
and clarify under which condition the linking of future and past null infinity presented in the next section is sensible and feasible.

The functions $\Theta$ and $\Xi$ transform, as can be seen from~\cref{eq:finBMS},
in the coadjoint representation of the centrally extended BMS group.
The phase space splits into disjoint orbits of the BMS group.
These orbits were classified in~\cite{Barnich:2015uva}; for a thorough introduction to the topic, consult~\cite{Oblak:2016eij}.
All solutions with different constant $\Theta$ or $\Xi$ belong to separate orbits, which means that these orbits can be uniquely labeled by their constant representative.
Relevant to the discussion are two additional families of orbits that do not admit constant representatives:
First, there is a two-parameter family of orbits with $\Theta = - 1$, but nonconstant $\Xi$.
Second, there are particular orbits without constant $\Theta$ representative, so called ``massless deformation'' orbits~\cite{Barnich:2014zoa}.
All other orbits do not have an energy bounded from below~\cite{Barnich:2014zoa}.
Positivity of the energy is a physically reasonable requirement, so these orbits are not considered in the following.

\begin{figure}
\centering
\includegraphics{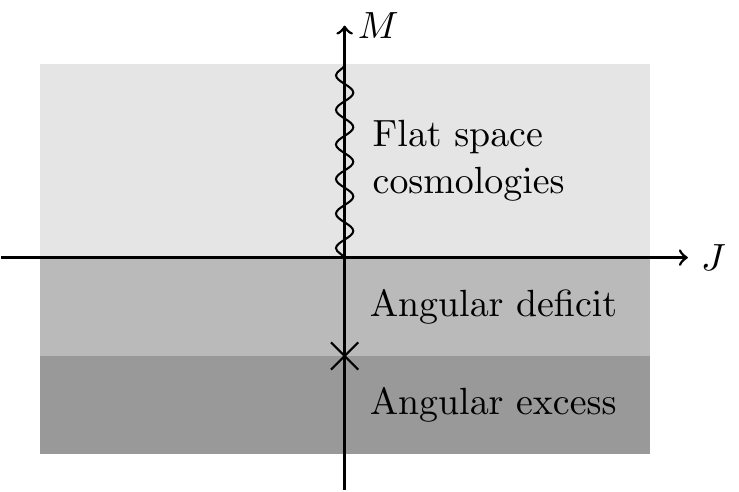}
\caption{The phase space of the spacetimes given in equation~\eqref{eq:constrep}.
The cross at $M=-1, J=0$ is Minkowski space.
The snake line indicates that the linking between past and future null infinity appears nonsensical at $M \ge 0, J = 0$.
The energy of a spacetime with angular excess is not bounded from below when acted upon by BMS transformations.}
\label{fig:MJ}
\end{figure}
\begin{figure}
\centering
\includegraphics{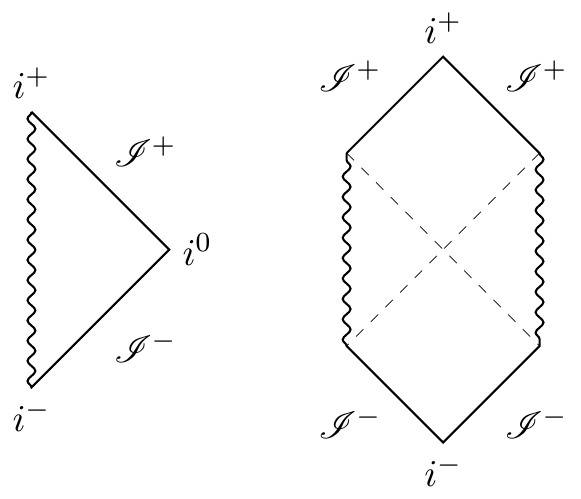}
\caption{Penrose diagrams for spacetimes with $M < 0$ (except $M=-1, J=0$ where there is no singularity) as well as spacetimes with $M = 0, J \neq 0$ (left) and flat space cosmologies (right).}
\label{fig:penrose}
\end{figure}

We take a closer look at orbits with constant representatives $\Theta^+(\phi) = M$ and $\Xi^+(\phi) = J/2$, summarized in \cref{fig:MJ}.
Here, $M$ and $J$ are, up to a factor~\footnote{
The factor is introduced to avoid clutter.
To recover true mass and angular momentum, use $M = 8 G M_{\mathrm{true}}$ and $J = 8 G J_{\mathrm{true}}$.
}, mass and angular momentum given by the charges~\cref{eq:charges}.
Then, at $\scri^+$ the metric is
\begin{align}
\label{eq:constrep}
  \dd s^2
  &= M \dd u^2 - 2 \dd u \dd r + J \dd u \dd \phi + r^2 \dd \phi^2
\end{align}
and similarly at $\scri^{-}$.
For strictly positive $M$ and nonvanishing $J$ the metric describes shifted boost orbi\-folds~\cite{Cornalba:2002fi,Cornalba:2003kd} which are quotients of Minkowski space.
They are also called flat space cosmologies and describe contracting and expanding phases separated by a region behind a cosmological horizon, see \cref{fig:penrose}.
They furthermore arise as a limit~\cite{Cornalba:2002fi} of Bañados-Teitelboim-Zanelli (BTZ) black holes~\cite{Banados:1992wn,Banados:1992gq}.
For vanishing $J$, we arrive at the boost orbifold~\cite{Khoury:2001bz,Seiberg:2002hr} with drastic changes in the geometric structure.
The spacetime where both $M$ and $J$ vanish is called the null-boost orbifold~\cite{Simon:2002ma,Liu:2002kb}.
In the last two cases there is a singularity between future and past infinity~\footnote{
See figure 5 and 9 in~\cite{Cornalba:2003kd}.},
so a mapping for $M \ge 0, J = 0$ seems unreasonable.
The ``O-plane''~\cite{Cornalba:2003kd} consists of orbits with $M = 0, J \neq 0$.

For strictly negative mass (left Penrose diagram in \cref{fig:penrose}) we distinguish between angular deficit ($-1 < M <0$) and angular excess ($M < -1$) solutions.
Minkowski space is at $M = -1$, $J = 0$.
While there are no black holes in three-dimensional flat space~\cite{Ida:2000jh},
angular deficit solutions describe point particles (rotating for nonvanishing $J$) and can be seen as the three-dimensional analog to Kerr metrics~\cite{Deser:1983tn,Deser:1983nh} (being axially symmetric vacuum solutions) or cosmic strings~\cite{Gott:1990zr} (see also \cite{Deser:1991ye,Cutler:1992nn}).

The linking of past and future null infinity presented in this paper is valid for all spacetimes that admit a constant representative, excluding $M \ge 0, J = 0$ (the snake line in \cref{fig:MJ}).
From the discussion above, we see that this includes nearly all physically relevant spacetimes, with the exception of the two-parameter family of orbits admitting $\Theta = -1$ as well as orbits where $\Theta$ belongs to the massless deformation.

\section{Linking Past and Future Null Infinity}
\label{sec:linking}

We now construct the map between $\scri^+$ and $\scri^-$ for spacetimes discussed in the previous section.
For this purpose we first introduce explicit coordinate systems.
One coordinate system will cover a neighborhood around $\scri^+$, the other one a neighborhood around $\scri^-$.
The map we then construct sends points at $\scri^+$ to points at $\scri^-$.
Since one coordinate system does not cover both of these regions, we describe the position of the point at $\scri^+$ in one coordinate system, and the position of the corresponding point at $\scri^-$ in the other coordinate system.

We consider spacetimes that admit a constant representative at $\scri^+$.
The first coordinate system $(u, r, \phi)$, that is introduced around $\scri^+$, is required to be such that the metric has the simple form~\eqref{eq:constrep}.
Notice that this coordinate system is defined only up to isometries of the spacetime.
Given this coordinate system we define the second coordinate system $(v, r, \phi')$ around $\scri^-$ by the following transformations.
\begin{align}
\shortintertext{$M > 0$, $J \neq 0$:}
&\begin{aligned}
  \label{eq:trafoGt0}
  u
  &= \frac{2 r}{M} + v
    - \frac{J}{2 M^{3/2}} \ln \left( 1 + \frac{4 r \sqrt{M}}{J - 2 r \sqrt{M}} \right)
  \\
  \phi
  &= \phi'
    + \frac{1}{\sqrt{M}} \ln \left( 1 + \frac{4 r \sqrt{M}}{J - 2 r \sqrt{M}} \right)
\end{aligned} \\
\shortintertext{$M = 0, J \neq 0$:}
&\begin{aligned}
  \label{eq:trafoEq0}
  u
  &= - \frac{8 r^3}{3 J^2} + v
  \qquad
  \phi
  = \phi' + \frac{4 r}{J}
\end{aligned}\\
\shortintertext{$M < 0$:}
&\begin{aligned}
  \label{eq:trafoLt0}
  u
  &= \frac{2r}{M} + v - \frac{J}{(-M)^{3/2}} \arctan \left( \frac{J}{2 r \sqrt{-M}} \right)
  \\
  \phi
  &= \phi' - \frac{2}{\sqrt{-M}} \arctan \left( \frac{J}{2 r \sqrt{-M}} \right)
\end{aligned}
\end{align}
These coordinate transformations fulfill the requirement that the second coordinate system does indeed cover $\scri^-$ (see \cref{sec:trafo}).
Apart from that, the form of the coordinate transformations is of no fundamental importance for the argument and they are chosen such that following equations are particularly simple.

We have now constructed and related our two coordinate systems.
The first one is defined up to isometries.
The second one is uniquely fixed by~\cref{eq:trafoGt0,eq:trafoLt0,eq:trafoEq0} once the first one is fixed.
We now define how points at $\scri^+$ are sent to points at $\scri^-$.

We send a point $A$ using coordinates $(u, r, \phi)$ at $\scri^+$ to a point $B$ at $\scri^-$ using coordinates $(v, r, \phi')$.
Any such map can be written as~\footnote{
This is different to the coordinate transformations~\cref{eq:trafoGt0,eq:trafoEq0,eq:trafoLt0}.
Plugging a point $P$ with the coordinates $(u_{P},r_{P},\phi_{P})$ into the transformations~\cref{eq:trafoGt0,eq:trafoEq0,eq:trafoLt0} leads to the \emph{same} point just in other coordinates $(v_{P},r_{P},\phi'_{P})$.}
  \begin{align}
    \label{eq:matching1}
    \begin{aligned}
      v_B &= f_1(u_A, \phi_A)
      \\
      \phi'_B &= f_2(u_A, \phi_A)
      \\
      r_B &= r_A = \infty \,,
    \end{aligned}
  \end{align}
with some functions $f_1$ and $f_2$.
Since the coordinate system $(u, r, \phi)$ is defined only up to isometries, one has to demand that the outcome of the mapping is independent of any such choice.
All spacetimes under consideration admit at least two isometries: Time translations, and rotations.
Time translations act as $u \to u + a$, and by~\cref{eq:trafoGt0,eq:trafoLt0,eq:trafoEq0}, also as $v \to v + a$.
Similarly, rotations act as $\phi \to \phi + b$ and $\phi' \to \phi' + b$.
Invariance under these isometries leads to the requirements that
\begin{align}
  \begin{aligned}
    f_1(u, \phi) + a &= f_1(u + a, \phi + b)
    \\
    f_2(u, \phi) + b &= f_2(u + a, \phi + b) \,,
  \end{aligned}
\end{align}
for all real numbers $a$ and $b$.
This almost fixes $f_1$ and $f_2$ and we find the invertible map
\begin{align}
\label{eq:matching2}
\begin{aligned}
  v_B
  &= u_A + c_1
  \\
  \phi'_B
  &= \phi_A + c_2
    \,,
\end{aligned}
\end{align}
with some constants $c_1$ and $c_2$.
The only invariant maps between $\scri^+$ and $\scri^-$ are of this form.

Now we fix the solely remaining freedom in our map, the constants $c_1$ and $c_2$.
To do this we consider Lorentz boosts on Minkowski space.
A Lorentz boost that is generated by a vector field~\footnote{
In Cartesian coordinates, the boost is generated by the vector field $t \pd_x + x \pd_t$, where $t = u + r = v - r$ and $x = r \cos \phi$.
} $- u \cos \phi \pd_u - \sin \phi \pd_\phi$ at $\scri^+$ is generated by $v \cos \phi' \pd_v + \sin \phi' \pd_{\phi'}$ at $\scri^-$.
The map~\eqref{eq:matching2} is invariant under this boost if and only if $c_1 = 0$ and $c_2 = \pi$.
Considering any other boost leads to the same conclusion.
We find that Minkowski space admits a unique invariant map.
We take $c_1$ and $c_2$ to be independent~\footnote{
This does not follow from our symmetry considerations and is the only choice in the derivation.}
of $M$ and $J$,
and arrive at the mapping prescription for spacetimes admitting constant representatives:
\begin{align}
\label{eq:matching3}
\begin{aligned}
  v_B
  &= u_A
  \\
  \phi'_B
  &= \phi_A + \pi
    \,.
\end{aligned}
\end{align}
Using symmetry arguments we found an antipodal relation in the angular coordinate as in the four-dimensional case~\cite{Strominger:2013jfa}.
Everything else falls into place.
A finite BMS transformation, parametrized by $\alpha$ and $f$, that acts on $\scri^+$ as
\begin{align}
\label{eq:BMS+m}
  \begin{aligned}
    u &\to u f'(\phi) + \alpha(f(\phi))
    \\
    \phi &\to f(\phi) \,,
  \end{aligned}
\end{align}
has to act with the same functions $\alpha$ and $f$ on $\scri^-$ as
\begin{align}
\label{eq:BMS-m}
  \begin{aligned}
    v &\to v f'(\phi' - \pi) + \alpha(f(\phi' - \pi))
    \\
    \phi' &\to f(\phi' - \pi) + \pi \,.
  \end{aligned}
\end{align}
This is the unique map between $\BMS^+$ and $\BMS^-$ that preserves the mapping~\cref{eq:matching3}.

Now we go back to the original goal of mapping asymptotic regions of spacetimes with any metric admitting a constant representative.
We take a metric that is given around $\scri^+$ as~\eqref{eq:3}.
By assumption we can apply a BMS transformation~\cref{eq:finBMS} to bring the metric into constant form~\cref{eq:constrep}.
Then we use the coordinate transformations~\cref{eq:trafoGt0,eq:trafoLt0,eq:trafoEq0} to find the metric around $\scri^-$
\begin{align}
  \dd s^2
  &= M \dd v^2 + 2 \dd v \dd r + J \dd v \dd \phi' + r^2 \dd \phi'^2
    \,.
\end{align}
Undoing the BMS transformation using the above relation between~\eqref{eq:BMS+m} and~\eqref{eq:BMS-m}, we finally get
a metric of the form~\eqref{eq:7}
with
\begin{align}
\label{eq:dofmatching}
  \begin{aligned}
    \Theta^+(\phi) &= \Theta^-(\phi + \pi)
    \\
    \Xi^+(\phi) &= \Xi^-(\phi + \pi) \,.
  \end{aligned}
\end{align}
From the definition of the charges~\cref{eq:charges} we immediately obtain infinitely many conservation laws,
\begin{align}
  Q^+_{T,Y}
  &= Q^-_{\tilde T, \tilde Y}
    \,,
\end{align}
one for every function $T(\phi) = \tilde T(\phi + \pi)$ and $Y(\phi) = \tilde Y(\phi + \pi)$.
The mapping is energy preserving: $Q^+_{1,0} = Q^-_{1,0}$.

\section{Adding Matter}
\label{sec:matter}

Up until now we have restricted ourselves to the vacuum solutions~\eqref{eq:3} and~\eqref{eq:7}.
Here we turn to the classical scattering problem of a massless field coupled to gravity, where initial and final data are prescribed on $\scri^-$ and $\scri^+$.
Both sets of data transform under each BMS group separately.
When considering BMS as a symmetry of the scattering problem, the separate symmetries of $\scri^+$ and $\scri^-$ must be broken to a single one.
Using the results of the vacuum case presented above, a similar mapping of symmetries can be achieved in the presence of matter, as follows.

We require that the solution to the Einstein equations admits some well defined spacelike infinity $i^0$ and that there is vacuum in a neighborhood of $i^0$.
Thus in this neighborhood around $i^0$, the metric will have the form~\cref{eq:3} and~\cref{eq:7}.
Using the algorithm established above we can find a mapping between $\scri^+$ and $\scri^-$, and consequently a relation between the two respective symmetry groups $\BMS^+$ and $\BMS^-$ according to~\eqref{eq:BMS+m} and~\eqref{eq:BMS-m}.
This mapping is a priori valid only in the neighborhood of $i^0$, in which the coordinate system~\eqref{eq:constrep} is well defined.
However, a BMS-transformation is determined on the entirety of $\scri^{\pm}$ by prescribing it on one cross section~\cite{Geroch:1977jn}.
The linking of $\BMS^+$ and $\BMS^-$ near $i^0$ is therefore enough to establish a linking on the whole of $\scri$, thus breaking the symmetry $\BMS^+\otimes\BMS^-$ to a single $\BMS$ acting on both $\scri^+$ and $\scri^-$.
In particular, the mapping~\eqref{eq:dofmatching} of the gravitational degrees of freedom near $i^0$ is still valid.
Given the flux of matter through $\scri^{\pm}$, these relations can be used as initial conditions for integrating the constraint equations along $\scri^\pm$, thus providing initial or final data for the scattering problem.

\section{Discussion}
\label{sec:discussion}

For three-dimensional spacetimes that admit a constant representative (see \cref{fig:MJ}) the map given by~\cref{eq:matching3} together with~\cref{eq:trafoGt0,eq:trafoLt0,eq:trafoEq0} provides a linking between future and past null infinity and their respective symmetry groups.
An immediate consequence of this linking is the existence of an infinite number of conservation laws, expressed in~\eqref{eq:dofmatching}.
This is just conservation of energy and angular momentum at every angle.

In the context of flat space holography, the two functions $\Theta$ and $\Xi$ can be seen as components of the stress-tensor of the dual boundary theory~\cite{Barnich:2012aw,Barnich:2013yka,Carlip:2016lnw}.
Due to the matching presented in this paper the two boundary theories defined on $\scri^+$ and $\scri^-$ are connected. It would be interesting to employ these relations by calculating boundary observables such as entanglement entropy.

The single $\BMS$ group, that was obtained from the linking, should be regarded as a symmetry for the S-matrix of three-dimensional Einstein gravity coupled to matter.
Further study is required to determine to what extent the relations between BMS symmetry, memory effect and soft theorems present in four dimensions~\cite{Strominger:2013jfa,He:2014laa,Strominger:2014pwa} are realized in three dimensions.

\ifaps\ifwordcount
\end{document}
\fi\fi

\begin{acknowledgments}
\uchyph=0
We thank Steve Carlip, Hernán González, Daniel Grumiller, Maria Irakleidou, Wout Merbis, Blagoje Oblak, Jan Rosseel, and Cédric Troessaert for valuable comments.
The authors acknowledge the scientific atmosphere at the Wörthersee Flat Space Workshop (FSF) where this project was initiated.
S.P., J.S. and F.S. were supported by the Austrian Science Fund (FWF) projects P~27396-N27, P~28751-N27 and P~27182-N27.
\end{acknowledgments}

\appendix

\section{Coordinate Transformations}
\label{sec:trafo}

The coordinate transformations \cref{eq:trafoGt0,eq:trafoEq0,eq:trafoLt0} are constructed such that the coordinates $(u, r, \phi)$ cover $\scri^+$, while $(v, r, \phi')$ cover $\scri^-$.
That this is true can most easily be seen for Minkowski space ($M = -1, J = 0$).
Here, $u = t - r$ and $v = t + r$ are usual retarded and advanced times.
Depending on which one is held fixed, one ends up at either $\scri^+$ or $\scri^-$ as $r$ goes to infinity.
On other spacetimes with $M \neq 0$ this works analogously.
We now discuss the more complicated case of flat space cosmologies ($M > 0$, $J \neq 0$).

\begin{figure}
\centering
\includegraphics{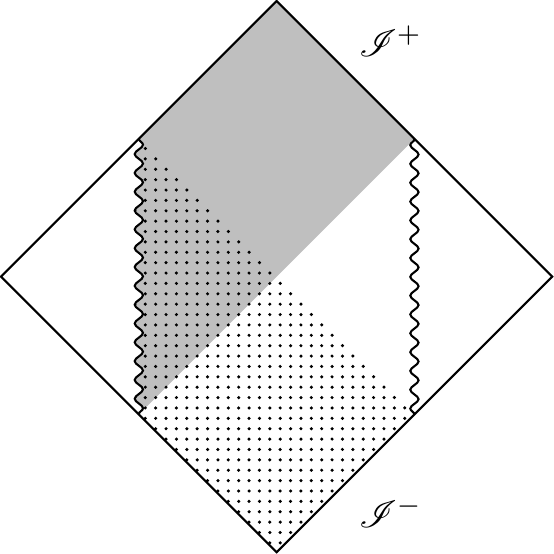}
\caption{Penrose diagram of a constant $Y$ slice of Minkowski space.
  The snake lines indicate where causal singularities develop when taking the quotient to obtain flat space cosmologies.
  The gray and the dotted regions mark different coordinate patches.}
\label{fig:fsc-quotient}
\end{figure}

Flat space cosmologies can be constructed as quotients of Minkowski space.
We use Cartesian coordinates $(T, X, Y)$ and define the coordinates ($u, r, \phi$) with $r > 0$ by
\begin{align}
  \begin{aligned}
  T
  &= \frac{r}{\sqrt{M}} \cosh \left( \sqrt{M} \phi \right)
    - \frac{J}{2M} \sinh \left( \sqrt{M} \phi \right)
  \\
  X
  &= \frac{r}{\sqrt{M}} \sinh \left( \sqrt{M} \phi \right)
    - \frac{J}{2M} \cosh \left( \sqrt{M} \phi \right)
  \\
  Y
  &= \frac{1}{\sqrt{M}} \left(
    - r + M u + \frac{J \phi}{2}
    \right)
    \,.
  \end{aligned}
\end{align}
The coordinates ($u, r, \phi$) cover the region
\begin{align}
  \begin{aligned}
  - \sqrt{T^2 + \frac{J^2}{4M^2}} < X < T
  &\quad \text{if} \quad J > 0
  \\
  -T < X < \sqrt{T^2 + \frac{J^2}{4M^2}}
  &\quad \text{if} \quad J < 0
    \,,
  \end{aligned}
\end{align}
which, for $J > 0$, corresponds to the gray region in \cref{fig:fsc-quotient}.
The metric in these coordinates is
\begin{align}
  \dd s^2
  &= M \dd u^2 - 2 \dd u \dd r + J \dd u \dd \phi + r^2 \dd \phi^2
    \,.
\end{align}
Upon identifying
\begin{align}
  \phi
  &\sim \phi + 2 \pi
\end{align}
we end up with flat space cosmologies parametrized by $M$ and $J$.
The identifications are given in Cartesian coordinates as
\begin{align}
  \begin{pmatrix}
    T \\ X \\ Y
  \end{pmatrix}
  &\sim
  \begin{pmatrix}
    T \cosh( 2 \pi \sqrt{M} ) + X \sinh( 2 \pi \sqrt{M} ) \\
    X \cosh( 2 \pi \sqrt{M} ) + T \sinh( 2 \pi \sqrt{M} ) \\
    Y + \frac{\pi J}{\sqrt{M}}
  \end{pmatrix}
  \,,
\end{align}
corresponding to a boost in $X$ direction plus a translation in $Y$ direction.
This is why flat space cosmologies are also referred to as shifted boost orbifolds~\cite{Cornalba:2002fi,Cornalba:2003kd}.
At $r=0$, where $X^2 - T^2 = {\left(\frac{J}{2M}\right)}^2$, null-like separated points become identified, leading to a causal singularity there.

A similar coordinate system ($v, r, \phi'$) can be defined as
\begin{align}
  \begin{aligned}
  T
  &= - \frac{r}{\sqrt{M}} \cosh \left( \sqrt{M} \phi' \right)
    - \frac{J}{2M} \sinh \left( \sqrt{M} \phi' \right)
  \\
  X
  &= - \frac{r}{\sqrt{M}} \sinh \left( \sqrt{M} \phi' \right)
    - \frac{J}{2M} \cosh \left( \sqrt{M} \phi' \right)
  \\
  Y
  &= \frac{1}{\sqrt{M}} \left(
    - r - M v - \frac{J \phi'}{2}
    \right)
    \,,
  \end{aligned}
\end{align}
carefully chosen such that the identifications $\phi' \sim \phi' + 2 \pi$ correspond to the ones before.
This coordinate system covers the dotted region in \cref{fig:fsc-quotient}.
The metric becomes
\begin{align}
  \dd s^2
  &= M \dd v^2 + 2 \dd v \dd r + J \dd v \dd \phi' + r^2 \dd \phi'^2
    \,.
\end{align}
In the region where the two coordinate systems overlap, we find the coordinate transformation given by \cref{eq:trafoGt0}.

\ifaps
\vspace{2em}
\fi

\bibliographystyle{utphys}
\bibliography{bibl}

\end{document}